# Further Considerations Regarding the Aharonov-Bohm Effect and the Wavefunction of the Entire System

**Allan Walstad**
Department of Physics, University of Pittsburgh at Johnstown, PA 15904
USA   email: awalstad@pitt.edu

**Abstract**  In an earlier paper it was demonstrated that the hypothesized electrostatic version of the Aharonov-Bohm ("AB") effect does not exist. The conclusion follows straightforwardly once one recognizes that interference takes place in the configuration space of the entire system, including the experimental apparatus, and the wavefunction of the apparatus cannot be ignored.  Two additional results are presented here.  1. Observations of interference that had been attributed to an analogue of the electrostatic AB effect (or "scalar effect") are actually due to a magnetic AB effect.  2. In the original magnetic AB effect itself, there is no phase shift if it is possible effectively to shield the solenoid from the influence of the passing electron.  This result is not in conflict with the landmark experiments of Tonomura and colleagues if Wang's recent claim is correct, that superconductive shielding could not have isolated the toroidal magnet from the magnetic pulse of the passing electron.

## 1 Introduction

In 1959 Aharonov and Bohm [1] noted that the effect of electromagnetic potentials on the wavefunction of a charged particle is to introduce a phase along any given spacetime path:

$$\varphi = (q/\hbar) \left[ \int V \, dt - \int \mathbf{A} \cdot d\mathbf{r} \right] \qquad (1)$$

where $V$ and $\mathbf{A}$ are the scalar and vector potentials, respectively.  In an interference experiment, if either $V$ or $\mathbf{A}$ differs along two interfering paths of the charged particle, the resulting phase difference would be expected to have an observable affect on the interference pattern.  Aharonov and Bohm described how this physical effect might be brought about in the absence of electric or magnetic fields and forces acting on the particle, thereby establishing an independent physical significance for the potentials, which had been generally regarded as mathematical tools for the calculation of fields. The hypothesized phenomenon became known as the Aharonov-Bohm ("AB") effect. The electrostatic version is associated with a difference in *V*, the magnetic version with a difference in **A**.

Clearly, the phase introduced by V is to be understood in terms of the electrostatic potential energy $\Delta E = qV$ which, acting for a time $\Delta t$, would generate a phase of magnitude $\Delta E \, \Delta t / \hbar = qV \, \Delta t / \hbar$. [2] Similarly, the phase introduced by **A** is due to the electromagnetic momentum $\Delta \mathbf{p} = q\mathbf{A}$ which, acting through a displacement $\Delta \mathbf{r}$, would generate a phase difference of magnitude $\Delta \mathbf{p} \cdot \Delta \mathbf{r} / \hbar = (q/\hbar) \, \mathbf{A} \cdot \Delta \mathbf{r}$.

As has recently been pointed out [3], however, interference takes place in the configuration space of the entire system, including the source of potentials acting on the particle. The relevant phase is not only the phase generated by the energy and momentum of the particle; if the phase associated with the rest of the system is sufficiently affected by its interaction with the particle, then it cannot be ignored. (In a 1961 paper [4], Aharonov and Bohm sought to demonstrate that their results would not be affected by considering the wavefunction of the entire system; the error in their treatment has been identified [3].) Vaidman [5], Wang [6], and McGregor et. al. [7] have also recognized this point. With this understanding, the non-existence of the *electrostatic* AB effect is obvious from conservation of energy. If the particle's energy changes by $\Delta E$, the energy of the rest of the system changes by $-\Delta E$. The total energy being independent of the path of the particle, the total phase generated by energy will be independent as well.

(It is in precisely this manner that conservation of energy underlies the principle of Maupertuis in classical mechanics, which permits use of a reduced Lagrangian, involving only momenta, to calculate the trajectory of a system in configuration space. The trajectory of least action is that for which there is maximum constructive interference with the quantum amplitudes for closely neighboring trajectories *of the entire system*. Conservation of energy ensures that the phase generated by energy is the same on all accessible trajectories.)

In the *magnetic* AB effect as originally described, the interfering particle paths traverse opposite sides of a long solenoid (Fig. 1). The azimuthal vector potential of the solenoid imparts a different electromagnetic momentum to the particle on each path, thereby generating a different phase as the particle moves past the solenoid to the detector.

Conservation of momentum demands that the solenoid acquire an equal and opposite momentum to that acquired by the particle; nevertheless, as the phase generated by momentum is the product of momentum times displacement, and as the solenoid's mass is much greater than the mass of the particle, it does not undergo an appreciable displacement and thus does not contribute an appreciable phase. Unlike the electrostatic case, here the phase difference attributable to the particle is not compensated by the rest of the system: there is an observable magnetic effect on the interference pattern, as was confirmed long ago with interference experiments involving electrons.

The purpose of this paper is to present two additional results. In Sect. 2 it is shown that observations of interference involving neutrons that had been attributed to an analogue of the electrostatic AB effect (or "scalar effect") are actually due to a magnetic AB effect. In Sect. 3 it is pointed out that in the original magnetic AB effect itself, if the solenoid can be shielded from the magnetic pulse of the passing electron, there is no observable AB phase shift: the phase of the solenoidal vector potential acting on the electron is cancelled by the phase of the same vector potential acting on the induced screening current.

## 2 Electrostatic AB analogue with neutrons?

The electrostatic AB effect as originally proposed has never been experimentally tested at sufficient precision to confirm or refute it. Nevertheless, there are observations of interference involving neutrons that have been interpreted as validating the underlying and more general concept of a "scalar" AB effect, in which a difference in potential energy of the particle produces a phase shift. That interpretation will be shown to be mistaken, and a correct explanation of the observations will be presented.

A neutron, though uncharged, possesses a magnetic moment with $z$ component $\pm\mu$. In a magnetic field $B$, there is a magnetic potential energy of magnitude $\Delta E = \pm\mu B$. In an interference experiment, suppose that one of the two interfering paths goes through a long, narrow solenoid, as illustrated in Fig. 2. If a current is applied to the solenoid to generate a magnetic field for a time $\Delta t$ while the neutron is safely within the interference apparatus, the energy difference would appear to generate a phase difference of magnitude

$$\Delta\varphi = \Delta E\, \Delta t / \hbar = \mu B\, \Delta t / \hbar \qquad (2)$$

between the two paths. A phase difference of just this magnitude has been observed experimentally [8, 9]. Since the uniform magnetic field in the solenoid exerts no force on the neutron, one might suppose that this result confirms the existence of a scalar effect quite analogous to the electrostatic AB proposal.

Nevertheless, this interpretation of the experimental results cannot be correct. The change in energy of the neutron comes at the expense of the current source for the solenoid. As total energy is conserved, independently of the path followed by the neutron, and as interference takes place in the configuration space of the entire system, the phase shift cannot be due to a change in energy.

The correct explanation is as follows. Associated with the magnetic moment of the neutron is an azimuthal vector potential (Fig. 3), which imparts an electromagnetic momentum to the conduction charges in the solenoidal windings. A phase is generated by the motion of the conduction charges constituting the current along this vector potential. Using a simplified model, it is easy to show that this phase is of the correct magnitude.

Let the $z$ axis be the axis of the solenoid, which has radius $R$. Consider a particle located on the $z$ axis having magnetic moment $\mu$ in the $\pm z$ direction. At distance $r$ in direction $\theta = \sin^{-1}(R/r)$, its vector potential, of magnitude $A = (\mu_0/4\pi)(\mu \sin\theta / r^2)$, points azimuthally along the solenoidal windings carrying current $I$. Let $N$ be the number of conduction charges, each with charge $q$ and moving at speed $v$, in *each turn of wire*. Then we have $I = Nqv/2\pi R$ and the contribution of this turn to the phase in time $\Delta t$ will be $\pm NqvA \Delta t/\hbar = \pm (2\pi RI)(\mu_0/4\pi)(\mu \sin\theta / r^2) \Delta t/\hbar$, depending on the orientation of the magnetic moment. Letting $n$ be the number of turns per meter along the $z$ axis, the contribution from the turns located in $dz$ will be

$$d\varphi = \pm (\mu_o/2) (\mu \sin\theta / r^2) R I n \, dz \, \Delta t / \hbar. \qquad (3)$$

Substituting $r = R / \sin\theta$ and $z = R / \tan\theta$ and integrating over $\theta$ we get a phase difference of magnitude

$$\Delta\varphi = \mu \, \mu_o \, n \, I \, \Delta t / \hbar = \mu \, B \, \Delta t / \hbar \qquad (4)$$

(compared with the field-free path), in agreement with (2).

## 3 The Magnetic AB Effect and Magnetic Shielding

In the theory of the magnetic AB effect as proposed by Aharonov and Bohm in 1959 and still widely accepted, the vector potential of a solenoid acts on the wavefunction of a charged particle to generate a phase difference between paths that traverse opposite sides of the solenoid. The predicted phase shift has long since been confirmed. Nevertheless, recognizing that interference takes place in the configuration space of the entire system, not just the particle, let us consider what effect the particle might have on the solenoid.

At a distance $r$, a particle of charge $q$ moving with velocity $\mathbf{v}$ produces a vector potential $\mathbf{A} = (\mu_o/4\pi)\, q\mathbf{v}/r$. Acting on the current $I$ in a directed element of wire $d\mathbf{s}$, in time $dt$ it generates a phase of magnitude

$$d\varphi = I\, \mathbf{A} \cdot d\mathbf{s}\, dt\, /\, \hbar$$

$$= I\, (\mu_o/4\pi)(q\, \mathbf{v}/r) \cdot d\mathbf{s}\, dt\, /\, \hbar. \tag{5}$$

On the other hand, the directed element of wire contributes $d\mathbf{A}' = (\mu_o/4\pi)\, I\, d\mathbf{s}/r$ to the vector potential at the location of the particle. The resulting contribution to the phase in time $dt$ is

$$d\varphi' = q\, d\mathbf{A}' \cdot \mathbf{v}\, dt\, /\, \hbar$$

$$= q\, (\mu_o/4\pi)(I\, d\mathbf{s}\, /\, r) \cdot \mathbf{v}\, dt\, /\, \hbar. \tag{6}$$

The expressions (5) and (6) for phase are identical. Either expression is correct, but their sum is not. Integrated over the solenoid and the time of flight of the particle, each yields the same correct result for the AB phase. (The interchangeability of the two methods of calculating the phase is demonstrated in a detailed analysis by McGregor et. al. [7] based on the Darwinian Lagrangian for the entire system.)

In the experiments of Tonomura and colleagues [10] that most securely confirmed the existence of the magnetic AB effect, a toroidal solenoid was isolated by a superconducting shield. The object was to rule out any possibility that the interfering electron paths were deflected by magnetic fields, but it has also been assumed that the solenoid itself would be shielded from the magnetic field of a passing charged particle. If the solenoid can be shielded from the influence of a charged particle, while the vector potential of the solenoid nevertheless acts on the particle, then equally valid methods of calculating the magnetic AB phase yield different results. Effective shielding would appear to leave us with a contradiction.

*The solution to this conundrum lies in the effect of the vector potential of the solenoid on the current that must be induced in the shielding in order to cancel the magnetic pulse from the passing charge.* The difference between the two interfering paths of the charged particle can be thought of in terms of an orbit of the particle around the solenoid, or, equivalently, in terms of a current that lasts for a short time. In the absence of shielding, the AB phase is the action of the solenoidal vector potential on this current. In order to shield the solenoid from the magnetic field of this current, an opposite current must be induced in the outer layer of the superconducting shield. It is obvious that the phase generated by the solenoidal vector potential acting on the induced current will tend to cancel the phase generated by the solenoidal vector potential acting on the passing charged particle.

That this cancellation is complete is also easy to see. The phase generated by the interaction of charges and currents via the vector potential is completely symmetric, as we saw with (5) and (6) above. Given the charge trajectories and currents, we can calculate the same phase two ways. In the absence of shielding, the phase arises *either* from the action of the passing electron on the currents responsible for the solenoidal magnetic field, *or* from the action of the solenoidal currents on the passing electron. Effective shielding requires that the induced shielding current acts on the solenoidal currents with an equal and opposite vector potential to that produced by the passing electron. By symmetry, the phase thereby produced is the same as what we would find from the solenoidal currents acting on the shielding current. Thus, there are two ways to calculate the phase shift in the presence of effective shielding: a) from the action of the passing particle and the induced current on the solenoidal currents, and b) from the action of the solenoidal currents on the induced current and the passing particle. Both ways give the same result, namely, zero. Therefore, if shielding is effective, there is no AB phase shift. (Note that the phase generated by the action of the particle on the induced shielding current is the same for both particle paths. Hence, it cancels out.)

This result corroborates the claim of Wang [11] that superconductive shielding could not have isolated the toroidal magnet from the effect of the passing electron. Wang's argument is that the timescale of the magnetic pulse from the electron corresponds to a frequency $f > kT_c/\hbar$, where $T_c$ is the superconducting critical temperature.

## 4 Concluding Remarks

If a quantum particle can reach a detector via two different paths, is the interference not fully determined by the relative phases *of the particle* on those paths? No, not if the particle path itself sufficiently affects phases in a larger system. We must take seriously that the arena of quantum interference is the many-dimensional configuration space of the entire system, even a macroscopic one. With this understanding, in a previous paper it was demonstrated that the electrostatic AB effect as originally proposed by Aharonov and Bohm does not exist. In the present paper, two further results are established. First, certain interference effects involving neutrons, which had been interpreted as an analogue of the electrostatic AB effect (a "scalar effect"), constitute instead a magnetic AB effect: the vector potential associated with the magnetic dipole of the neutron acts on a current in the apparatus to generate the observed phase shift. Second, in the magnetic AB effect itself as originally proposed, if it were possible effectively to shield the solenoid from the influence of the electron, there would be no phase shift.

**Acknowledgements** The author thanks Barry Torau and Nick Wiencek for technical assistance in preparing the manuscript.

**Figure Captions**

Fig. 1  The magnetic Aharonov-Bohm effect.  Wavepackets traverse opposite sides of a solenoid (viewed end-on).

Fig. 2  Magnetic phase shift with neutrons.  The current is turned on and off again, all while the wavepacket is within the interference apparatus.

Fig. 3  The vector potential of a neutron interacts with the current in a solenoid.

Fig. 4  Inward-moving magnetic field lines are cancelled by induced current at the surface of a superconducting cylinder.

Fig. 5  Magnetic field lines approaching from one side induce a current; a current front propagates through the superconducting cylinder to the far side.

Fig. 6  The current front propagates around the hollow interior, leaving an azimuthal current at the inner surface of the superconductor.

Fig. 1

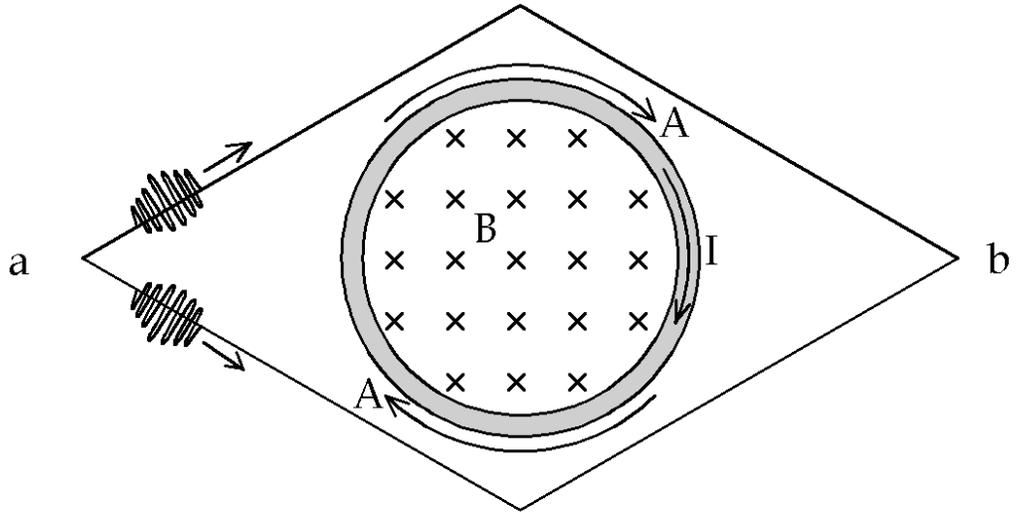

Fig. 2

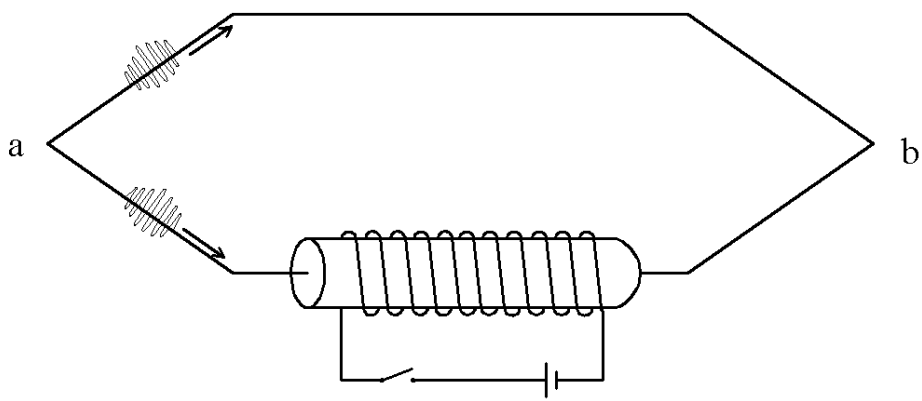

Fig. 3

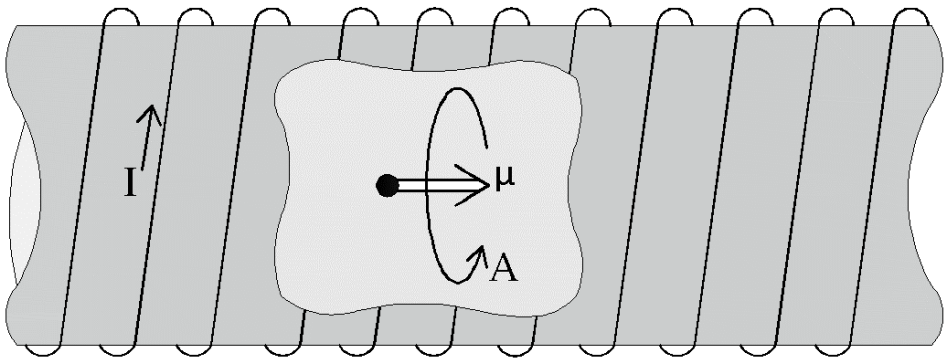